\documentclass[aps, twocolumn, showpacs]{revtex4}

\pdfoutput=1

\usepackage{amsmath}
\usepackage{amssymb}
\usepackage{graphicx}
\usepackage{xspace}

\newcommand{\eg}{{e.g.,\/}\xspace}
\newcommand{\ie}{{i.e.,\/}\xspace}

\newcommand{\eq}[1]{(\ref{#1})}
\newcommand{\Eq}[1]{Eq.~(\ref{#1})}
 
\newcommand{\Eqsc}[2]{Eqs.~(\ref{#1}), (\ref{#2})}
\newcommand{\Eqsd}[2]{Eqs.~(\ref{#1})-(\ref{#2})}
\newcommand{\Sec}[1]{Sec.~\ref{#1}}
\newcommand{\App}[1]{Appendix~\ref{#1}}
\newcommand{\Fig}[1]{Fig.~\ref{#1}} 
\newcommand{\Ref}[1]{Ref.~\cite{#1}}
\newcommand{\Refs}[1]{Refs.~\cite{#1}}

\newcommand{\mc}[1]{\mathcal{#1}}
\newcommand{\mcc}[1]{\mathfrak{#1}}

\newcommand{\favr}[1]{\langle #1\rangle}
\newcommand{\avr}[1]{\left\langle #1 \right\rangle}
\renewcommand{\vec}[1]{{\boldsymbol{\rm #1}}}
\newcommand{\unitvec}[1]{\hat{\vec{#1}}}

\newcommand{\dvectheta}{\dot{\vec{\theta}\kern 3pt}\kern -2.5pt}
\newcommand{\parallelind}{|\kern -1pt|}
\newcommand{\fullnabla}{\nabla_{*}}
\newcommand{\unitb}{\kern -2pt\unitvec{\kern 3pt b\kern 1pt}}
\newcommand{\avaprimesq}{\langle {a'}\kern .3pt^2\rangle}
\newcommand{\pprime}{{{p\kern 1pt}'{\kern -3pt}}}
\newcommand{\ppprime}{{{p\kern 1pt}_\perp'{\kern -3pt}}}

\newcommand{\supcomp}[1]{{\kern .5pt (\text{\raisebox{1pt}{$\scriptscriptstyle #1$}})}}
\newcommand{\supplus}{\supcomp{+}}
\newcommand{\supminus}{\supcomp{-}}
\newcommand{\suppm}{\supcomp{\pm}}
\newcommand{\suppar}{\supcomp{\parallelind}}
\newcommand{\subpar}{{\scriptscriptstyle \parallelind}}

\newcommand{\munu}{{\mu\kern -.5pt\nu}}
\newcommand{\fourvecm}{A'_\mu}
\newcommand{\fourvecl}{A_\mu}
\newcommand{\fbgl}{F_\munu}
\newcommand{\fbgm}{F'_\munu}
\newcommand{\fperp}{\tilde{F}_\munu}

\newcommand{\drift}[1]{\bar{#1}}
\newcommand{\dv}{\drift{v}}
\newcommand{\ppar}{{p\kern 1pt}_\subpar}
\newcommand{\dpp}{\drift{p}\kern .5pt}
\newcommand{\dvp}{\drift{\vec{p}}}
\newcommand{\dgamma}{\drift{\gamma}}
\newcommand{\dvv}{\drift{\vec{v}}}
\newcommand{\dvr}{\drift{\vec{r}}}
\newcommand{\meff}{m_\text{eff}}

\begin{document}

\title{Positive and negative effective mass of classical particles in oscillatory and static fields}
\author{I. Y. Dodin and N. J. Fisch}
\affiliation{Department of Astrophysical Sciences, Princeton University, Princeton, NJ 08544}
\date{\today}

\begin{abstract}
A classical particle oscillating in an arbitrary high-frequency or static field effectively exhibits a modified rest mass $\meff$ derived from the particle averaged Lagrangian. Relativistic ponderomotive and diamagnetic forces, as well as magnetic drifts, are obtained from the $\meff$ dependence on the guiding center location and velocity. The effective mass is not necessarily positive and can result in backward acceleration when an additional perturbation force is applied. As an example, adiabatic dynamics with $m_\subpar > 0$ and $m_\subpar < 0$ is demonstrated for a wave-driven particle along a dc magnetic field, $m_\subpar$ being the effective longitudinal mass derived from $\meff$. Multiple energy states are realized in this case, yielding up to three branches of $m_\subpar$ for a given magnetic moment and parallel velocity.
\end{abstract}



\maketitle

\section{Introduction}

A large number of problems connected with multiscale adiabatic dynamics of classical particles in oscillatory and static fields enjoy critical simplification within the guiding-center approach, which allows separating fast oscillatory motion of the particles from their slow translational motion \cite{book:northrop, ref:dewar73, ref:brizard07, ref:tao07, ref:qin07, ref:similon86, my:tunnel, my:dipole, foot:quant}. Often, the average forces on a guiding center are then written in terms of effective potentials $\Psi$, such as ponderomotive \cite{ref:gaponov58, ref:motz67, my:tunnel, my:invited}, diamagnetic \cite{book:jackson}, or others \cite{my:invited, my:dipole}. Yet the applicability of the potential approximation is limited to, at most, nonrelativistic interactions, another drawback being the unphysical difference in fictitious fields $-\nabla\Psi$ seen by different species.

The purpose of this paper is to offer an alternative approach that allows \textit{embedding} the average forces into the guiding center properties, through redefining the particle rest mass $\meff$. The average acceleration is then attributed to the effective mass variations, which are naturally different for different species; hence no fictitious fields are introduced. By definition, this ``object-oriented'' formulation \cite{foot:dressed} is also intrinsically relativistic. Therefore, it equally holds for arbitrary adiabatic interactions \cite{foot:gravity}, thus proving to be more fundamental as compared to the effective-potential approach.

Previously, the effective mass $\meff$ was similarly introduced for an electron interacting with an intense laser wave in vacuum, with additional fields considered only as perturbations \cite{my:meff, ref:kibble66, ref:eberly68, ref:sarachik70, ref:rax92, ref:rax92b, ref:rax93, ref:rax93b, ref:moore94, ref:bauer95, ref:mora97, ref:tokman99, my:gev, ref:bourdier05b}. In this paper, we show that $\meff$ can be defined as well for any other multiscale dynamics of a particle in high-frequency or static fields. We offer a general formula for the effective mass and show how manipulations of $\meff$ as a function of the guiding center variables yield the average forces and particle trajectories. We also show that the effective mass is not necessarily positive and can result in backward acceleration when an additional force is applied. As an example, we explore the average motion of a laser-driven particle immersed in a dc magnetic field. Multiple energy states are realized in this case and yield up to three branches of $\meff$ and the effective longitudinal mass $m_\subpar$ for a given magnetic moment and parallel velocity. We show that both $m_\subpar > 0$ and $m_\subpar < 0$ are possible then, the negative-mass regime too allowing for adiabatic dynamics.

The paper is organized as follows. In \Sec{sec:emass}, we derive the general formula for $\meff$ and the guiding center Hamiltonian accounting for additional perturbation fields, if any. In \Sec{sec:bdc}, we apply the effective mass formalism to the particle motion in a static magnetic field and rederive both the particle Hamiltonian and the magnetic drifts. In \Sec{sec:wave}, we explore the average motion of a laser-driven particle in a static magnetic field and demonstrate the possibility for adiabatic dynamics at negative $\meff$ and $m_\subpar$. In \Sec{sec:conc}, we summarize our main ideas. Supplementary calculations are given in Appendixes: In \App{sec:dlagr}, we obtain the general form of the drift Lagrangian employed in \Sec{sec:emass}. In \App{sec:ponder}, we show how the effective mass formalism allows derivation of ponderomotive forces in various cases of interest.

\section{Effective mass}
\label{sec:emass}

Consider a particle undergoing arbitrary quasi-periodic oscillations superimposed on the average motion. In the adiabatic regime, one can map out the quiver dynamics by changing variables \cite{book:northrop, ref:dewar73, ref:brizard07, ref:tao07, ref:qin07, ref:similon86, my:tunnel, my:dipole}; hence the guiding center is treated as a ``dressed'', or quasi-particle. Suppose, for now, that the background fields causing the oscillations do not vary along the trajectory. Then the associated field tensor $\fbgl$ will not enter the averaged equations as a force. However, it will affect the motion such that, in response to \textit{additional} perturbation fields $\fperp$, the guiding center will react as if it had a modified mass.

The effect is shown as follows. At zero $\fperp$, the average dynamics is determined only by the field tensor $\fbgm$ seen by the particle in the guiding-center rest frame $K'$, further denoted by prime. The action increment $d\mc{S}=\mc{L}\,dt$ in the laboratory frame $K$ is then written as
\begin{equation} 
\label{eq:action8}
d\mc{S}(\fourvecl,\dvv)= d\mc{S}'(\fbgm) + d\mc{G},
\end{equation}
where $\fourvecl$ is the four-potential such that $\fbgl = \partial_\mu\kern -.7pt A_\nu - \partial_\nu \kern -.7pt A_\mu$, $\dvv= \avr{\vec{v}}$ is the guiding center velocity in $K$, $\vec{v}$ is the particle true velocity, $\avr{\ldots}$ denotes the time-average, and $d\mc{G}$ depends on the selected gauge. Use $dt=\dgamma\, dt'$, where $dt'$ is the time interval in $K'$, $\dgamma=(1-\dv^2/c^2)^{-1/2}$, and $c$ is the speed of light. Then the guiding center Lagrangian reads
\begin{equation} 
\label{eq:lagr31}
\mc{L}=\mc{L}'/\dgamma + \dot{\mc{G}},
\end{equation}
where $\mc{L}'=d\mc{S}'/dt'$. Omitting an insignificant full time derivative, one can rewrite \Eq{eq:lagr31} as \cite{foot:jackson}
\begin{equation} 
\label{eq:lagr4}
\mc{L}=-\meff c^2\sqrt{1-\dv^2/c^2}.
\end{equation}
Hence $\mc{L}$ is formally equivalent to the Lagrangian of a free particle with an effective mass
\begin{equation} 
\label{eq:meff1}
\meff=-\mc{L}'/c^2
\end{equation}
that is, by definition, both gauge- and Lorentz-invariant.

By definition, $\mc{L}'=\mc{L}'(\fbgm)$, where $\fbgm$ can be written in terms of $\fourvecm$; thus
\begin{equation} 
\label{eq:lagr32}
\mc{L}'_{\fourvecm}=\frac{\mc{L}_{\fourvecl}-\dot{\mc{G}}_{\fourvecl}} {\sqrt{1-\dv^2/c^2}},
\end{equation}
for any $\dvv$. Consider $\dv \to 0$; in this case, $\fourvecl \to \fourvecm$, so
\begin{equation} 
\label{eq:dg}
\mc{L}'(\fbgm)=\big[\mc{L}_{\fourvecm}-(\partial G_{\fourvecm}/\partial t)\big]_{\dv = 0},
\end{equation}
where we removed the subindex $``\fourvecm$'' in the left-hand side, as $\mc{L}'$ is gauge-invariant.

In the absence of oscillations, $\mc{L}'$ must equal $-mc^2$, where $m$ is the true mass; therefore,
\begin{equation} 
\label{eq:lagr35}
\mc{L}'=-mc^2+\mc{L}_{\fourvecm}(\dv=0)-\mc{L}_{\fourvecm}(v=0).
\end{equation}
For clarity, we assume below that $L(v=0)=-mc^2$. Then, using \Eq{eq:lagr1}, one can write $\meff$ as
\begin{equation} 
\label{eq:meff}
\meff=\frac{1}{c^2}\big(\,\vec{J}\cdot\vec{\nu}-\avr{L}\big)_{\dv=0},
\end{equation}
where the right-hand side is to be evaluated in $K'$ (hence the index ``\,$\dv = 0$''); $\vec{J}$ are the actions and $\vec{\nu}$ are the frequencies of oscillations in canonical angles, if any, to average over (\App{sec:dlagr}). Therefore, apart from the latter, \textit{$\meff$ is proportional to the gauge-independent part of the averaged Lagrangian in the guiding-center rest frame} \cite{foot:routhian}. Since $\mc{L}'$ is calculated in $K'(\dvv)$, $\meff$ is generally a function of the velocity $\dvv$. When $\fbgm$ slowly varies with the guiding center coordinate $\dvr$ or time $t$, $\meff$ may similarly depend on those as well, so \Eq{eq:lagr4} will automatically yield the average forces [\Eq{eq:euler}].

Suppose now that a particle interacts with a perturbation field $\fperp$ governed by $\tilde{A}_\mu=(\tilde{\vec{A}},\tilde{\varphi})$, which is imposed over $\fbgl$ \cite{ref:moore94, my:meff, ref:khazanov02}. In the adiabatic regime, the orbit is not altered on the oscillation time scale; thus,
\begin{equation} 
\label{eq:lagr5}
\mc{L}=-\meff c^2\sqrt{1-\dv^2/c^2}+\frac{e}{c}(\dvv\cdot\tilde{\vec{A}})-e\tilde{\varphi}
\end{equation}
($e$ being the particle charge), and a nonelectromagnetic potential can be added similarly. Then, the canonical momentum equals $\drift{\vec{P}}=\drift{\vec{p}} +(e/c)\tilde{\vec{A}}$, and the kinetic momentum $\drift{\vec{p}}$ is given by
\begin{equation} 
\label{eq:moment1}
\dvp = \dgamma m_{\rm eff}\dvv-\frac{c^2}{\dgamma}\,\frac{\partial \meff}{\partial \dvv}.
\end{equation}
Correspondingly, the Hamiltonian $\mc{H}=\drift{\vec{P}}\cdot \dvv-\mc{L}$ reads
\begin{equation} 
\label{eq:qenergy}
\mc{H}=\dgamma \meff c^2-\frac{c^2}{\dgamma}\,\left(\dvv\cdot\frac{\partial \meff}{\partial \dvv}\right)+e\tilde{\varphi},
\end{equation}
and $\mc{E}=\mc{H}(\dvr,\drift{\vec{P}},t)$ is conserved when $\mc{H}$ is independent of time. Thus, $\meff$ can be viewed also as the normalized quasi-energy of an unperturbed ($\fperp=0$) particle in the guiding-center rest frame, $\meff=\mc{E}'\kern -1pt/c^2$.

\section{Static magnetic field}
\label{sec:bdc}

Let us demonstrate how the effective mass formalism applies to the problem of particle motion in a dc magnetic field $\vec{B} = \nabla \times \vec{A}$, where
\begin{equation} 
\label{eq:lagrtrue}
L=-\frac{mc^2}{\gamma}+\frac{e}{c}(\vec{v}\cdot \vec{A}),
\end{equation}
and $\gamma=(1-v^2/c^2)^{-1/2}$. Assuming a smooth $\vec{B}$, the motion can be averaged over Larmor oscillations at frequency $\Omega=eB/mc\gamma$, so the guiding center dynamics that remains is one-dimensional, and the associated action $J = (mc/e)\,\mu$ is conserved \cite{book:jackson}, where $\mu = p^{\kern 1 pt 2}_\perp/2Bm$ is the magnetic moment, and $\vec{p}_\perp=\gamma m\vec{v}_\perp$ is the relativistic kinetic momentum transverse to $\vec{B}$. Thus, the effective mass reads $\meff =[\mu B/\gamma'-\avr{L}']/c^2$, where the prime denotes the guiding-center rest frame $K'$; $\mu$ and $B$ are Lorentz invariants, and
\begin{equation}
\gamma'=\sqrt{1+2\kern 1pt\mu\kern -.3pt B /mc^2}
\end{equation}
is constant. Since $\avr{L}' =  -(mc^2+\mu B)/\gamma'$, one obtains
\begin{equation} 
\label{eq:meffb}
\meff=m\sqrt{1+2\kern 1pt\mu\kern -.3pt B/mc^2},
\end{equation}
which is a relativistic invariant. The guiding center momentum is then given by $\dgamma \meff\dv=\ppar$, where we used a Lorentz transformation $\gamma=\dgamma\gamma'$. Hence the Hamiltonian~\eq{eq:qenergy} reads \cite{ref:littlejohn84, ref:boozer96}
\begin{equation}
\label{eq:hammu} 
\mc{H}=\sqrt{m^2c^4+2\mu\kern -.3pt B\kern 1pt mc^2 + \ppar^2 c^2}+e\tilde{\varphi},
\end{equation}
yielding, after omitting an insignificant constant, the well known nonrelativistic limit
\begin{equation}
\label{eq:hammu2} 
 \mc{H} = \frac{1}{2m}\,\ppar^2+\mu B + e\tilde{\varphi}.
\end{equation}

A more precise calculation also delivers particle drifts \cite{book:northrop, book:jackson, ref:boozer96}: Allow arbitrary $\dv_\subpar$, yet assume nonrelativistic $\dvv_\perp$ so as to treat the transverse drift as a perturbation. Following \Ref{ref:danilkin95}, we write the new guiding center Lagrangian as $\mc{L}=\mc{L}_0 + \mc{L}_\text{int}$, where $\mc{L}_0=-\meff c^2/\dgamma$ is that for the unperturbed motion, and 
\begin{equation} 
\mc{L}_\text{int}= \frac{e}{c}(\dvv\cdot\vec{A})-e\tilde{\varphi}
\end{equation}
is the interaction Lagrangian small compared to $\mc{L}_0$. For simplicity, assume static fields, so the guiding center quasi-energy~\eq{eq:qenergy} is conserved. In this case, we can consider $\dvv$ as a function of $\dvr$; hence $\delta \mc{S} = 0$ yields
\begin{equation} 
\label{eq:pdrift}
(\dvv \cdot\! \fullnabla)\,\dvp = \frac{e}{c}\,\dvv \times \vec{B}-\frac{c^2}{\dgamma}\,\nabla \meff,
\end{equation}
where $\nabla$ differentiates with respect to $\dvr$ at fixed $\dvv$, and $\fullnabla \equiv \nabla + \sum_i(\nabla \dv_i)(\partial/\partial \dv_i)$ is the full spatial derivative. \Eq{eq:pdrift} is equivalent to 
\begin{equation} 
\label{eq:vcrossb}
\dvv \times \vec{B}^* = 0, \quad 
\vec{B}^* = \vec{B} + (c/e)\,\fullnabla\! \times \dvp,
\end{equation}
which can as well be put in the form
\begin{equation} 
\label{eq:vdrift}
\dvv = 
\drift{v}_\subpar\,\frac{\vec{B} + (c/e)\,\fullnabla\!\times\!(\ppar \unitb)}{B+(c/e) \ppar\,\unitb\! \cdot\! (\nabla \!\times \unitb)},
\end{equation}
where $\unitb = \vec{B}/B$, and $\dvp \approx \ppar \unitb$. This generalizes a similar analysis, which was proposed in \Ref{ref:danilkin95} for $v \ll c$, to any $v$, such that $\dv_\perp \ll c$. 

\Eqsc{eq:vcrossb}{eq:vdrift} yield the known expressions of the traditional drift approximation \cite{ref:grebogi84, ref:boozer80, ref:white82, ref:boozer84, ref:littlejohn84, ref:tao07}. However, they also allow for an arbitrary $\meff$, not necessarily that given by \Eq{eq:meffb}; thus additional strong fields, if any, are as well embedded here. Derivation of time-dependent and fully relativistic magnetic drifts \cite{ref:pozzo98, ref:beklemishev99} using the effective mass formalism should be possible, too, but remains out of the scope of the present paper.

\section{Relativistic wave field over a static magnetic field}
\label{sec:wave}

The unified effective mass formulation readily yields the ponderomotive forces previously derived from other considerations (\App{sec:ponder}). In this section, we contemplate another example of particle ponderomotive dynamics, which exhibits unusual properties that, to our knowledge, have not been covered in literature.

\subsection{Basic equations}
\label{seq:wavebasic}

Consider a relativistic particle in a wave propagating along a static magnetic field \cite{ref:roberts64, ref:qian99, ref:qian00, ref:bourdier00, ref:salamin00, ref:komarov01, ref:ondarza01, ref:kong07b, ref:kong07}. Assume a smooth magnetic field $\vec{B}=\nabla \times \vec{A}_\text{dc}$, approximately in the $\unitvec{z}$ direction; then the vector potential $\vec{A}_\text{dc}$ can be considered a linear function of the particle displacement from the guiding center location:
\begin{equation}
\label{eq:ab}
\vec{A}_\text{dc} = \frac{1}{2}\,B(z)\,\big(\vec{\unitvec{z}\times\vec{r}_\perp}\big).
\end{equation}
For simplicity, we will also assume a vacuum electromagnetic wave with circular polarization in the plane transverse to $\vec{B}$, so the corresponding vector potential reads
\begin{equation}
\label{eq:aw}
\vec{A}_\text{w} =\left(\frac{mc^2}{e}\right) \frac{a_0}{\sqrt{2}}\,\big(\unitvec{x}\cos\xi - \unitvec{y}\sin \xi\big),
\end{equation}
where the invariant $a_0=eE_0/mc\omega$ is allowed to slowly vary in space and time, $E_0$ is the amplitude of the electric field $\vec{E} = -(1/c)(\partial \vec{A}_\text{w}/\partial t)$, and $\xi = \omega t-kz$ is the phase, with $k=\omega/c$. In this case, the particle motion is fully integrable, and the problem can be solved analytically.

According to \Sec{sec:emass}, we calculate $\meff$ for uniform fields, using \Eq{eq:lagrtrue} with $\vec{A}=\vec{A}_\text{dc}+\vec{A}_\text{w}$. The effective mass is determined by the averaged Lagrangian in the guiding-center rest frame $K'$ (further denoted by prime), which is found as follows. Since $\vec{A}$ depends on $z$ and $t$ only via $\xi(z,t)$, there exists an integral
\begin{equation}
\label{eq:ucons}
u=\gamma-p_z/mc,
\end{equation}
yielding that the following equality holds for any $\!f$ \cite{my:meff}:
\begin{equation}
\label{eq:avr}
\avr{f} = \avr{\gamma\kern -1pt f}_\xi\!/\!\!\avr{\gamma}_\xi.
\end{equation}
(The subindex $\xi$ denotes that the averaging is performed over the phase rather than time.) Take $f=L'$; then
\begin{equation}
\label{eq:avrellagr}
\avr{L}'=-\frac{1}{\avr{\gamma'}_\xi}\,\left(mc^2-\frac{e}{mc}\avr{\vec{p}'\cdot \vec{A}'}_\xi\right).
\end{equation}
With $f=\vec{v}$, \Eq{eq:avr} also yields
\begin{equation}
\label{eq:dvvrel}
\dvv = \avr{\vec{p}}_\xi\!/m\!\avr{\gamma}_\xi, \quad \avr{\vec{p}'}_\xi = 0, \quad \avr{\gamma'}_\xi = u'.
\end{equation}
Given that the average motion is solely in the $\unitvec{z}$ direction, $K'$ is now defined as the frame where $\avr{\pprime_z}_\xi = 0$. 

Hence the particle motion can be written as follows:
\begin{subequations}\label{eq:xyz}
\begin{align}
& x = x_0+\mc{R} \cos \theta -\frac{a_0}{ku\sqrt{2}}\,\frac{\sin \xi}{(1-\sigma)},\\
& y = y_0-\mc{R} \sin \theta  -\frac{a_0}{ku\sqrt{2}}\,\frac{\cos \xi}{(1-\sigma)},\\
& z = z_0+\frac{\rho_0\xi}{k u}+\frac{\mc{R}a_0}{u\sqrt{2}}\,\frac{ \sigma\cos (\xi-\theta)}{(1-\sigma)^2},\\
& p_x = - \mc{P} \sin \theta -\frac{mc a_0}{\sqrt{2}}\,\frac{\cos \xi}{(1-\sigma)},\\
& p_y = -\mc{P} \cos \theta  +\frac{mc a_0}{\sqrt{2}}\,\frac{\sin \xi}{(1-\sigma)},\\
& p_z = mc\rho_0-\frac{\mc{P}a_0}{u\sqrt{2}}\,\frac{\sin (\xi-\theta)}{(1-\sigma)}.
\end{align}
\end{subequations}
Here $\rho_0 = u\dv/(c-\dv)$ is the normalized phase-averaged momentum; $\mc{R}\equiv \mc{P}/m\Omega_0$, $\mc{P}$, and $\theta = \sigma \xi+\theta_0$ denote the gyroradius, the transverse momentum, and the phase of the \textit{free} gyromotion superimposed on the wave-induced oscillations, $\sigma$ being a Lorentz invariant:
\begin{equation}
\label{eq:sigma}
\sigma = \frac{\Omega_0}{\omega u} = \frac{\Omega_0/\gamma}{\omega - k v_z};
\end{equation}
$\mu\equiv\mc{P}^2/2Bm$ is the invariant action of this gyromotion conserved under adiabatic perturbations [cf. \Eq{eq:murf}]; $x_0$, $y_0$, $z_0$, $\dv$, $\mc{P}$, $\theta_0$ are determined by initial conditions. 

To find $u$, substitute $\gamma=u+\rho_z$ into $\gamma^2=1+\rho_\perp^2+\rho_z^2$, where $\vec{\rho}\equiv\vec{p}/mc$, and average over $\xi$ using $\avr{\rho_z}_\xi = \rho_0$:
\begin{equation}
\label{eq:up1}
(u + \rho_0)^2=1+\rho_0^2+\avr{\rho_\perp^2}_\xi.
\end{equation}
Then,
\begin{equation}
\label{eq:up}
u=h\,\sqrt{\,1+s^2+\frac{a^2_0}{2(1-\sigma)^2}}, \quad h=\sqrt{\frac{c-\dv}{c+\dv}},
\end{equation}
where $\sigma=\sigma(u)$, $s^2 \equiv 2\mu B/mc^2$ is an invariant, and $\frac{1}{4}mc^2a^2_0$ equals the zero-$B$ nonrelativistic ponderomotive potential $\Phi=e^2E^2_0/4m\omega^2$ [\Eq{eq:classphi}]. 

\begin{figure}
\centering
\includegraphics[width=0.4 \textwidth]{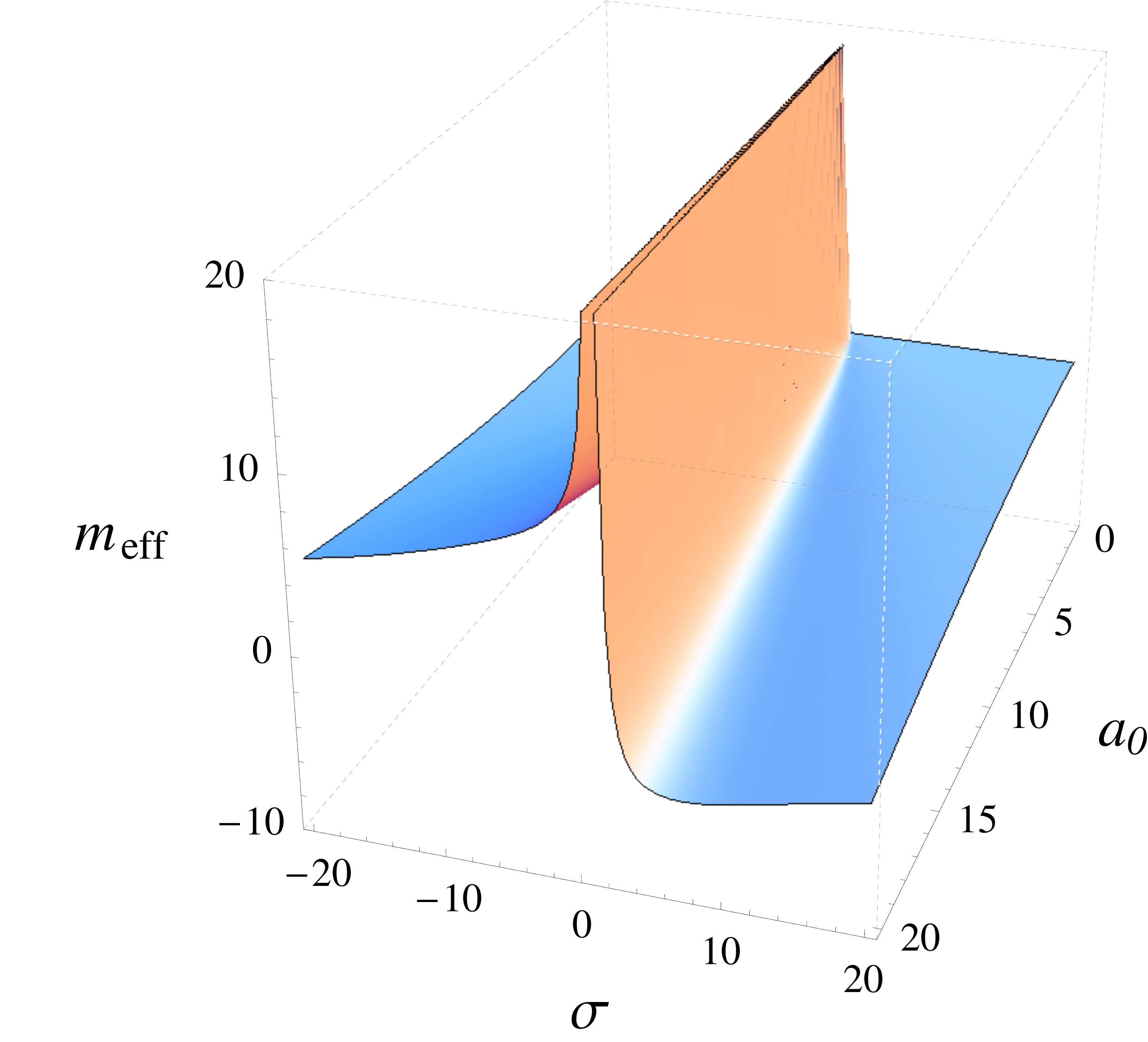}
\caption{(color online) Effective mass $\meff$ of a wave-driven particle in a magnetic field [\Eq{eq:meffw}; $m$ units] vs. $a_0$ and $\sigma$.}
\label{fig:meff}
\end{figure}

Combining \Eqsd{eq:avrellagr}{eq:up}, one gets
\begin{equation}
\label{eq:meffw}
\kern -2pt\frac{\meff}{m}\!=\!\left[1+s^2+\frac{a^2_0(2-\sigma)}{4(1-\sigma)^2}\right]\!\!\! \left[1+s^2+\frac{a^2_0}{2(1-\sigma)^2}\right]^{\!{\scriptscriptstyle -}\frac{1}{2}}\kern -12pt,
\end{equation}
which is a covariant form of $\meff$ for the effective mass is expressed as a function of Lorentz invariants. \Eq{eq:meffw} yields \Eqsd{eq:meffb}{eq:vdrift} at $a_0=0$, \Eqsc{eq:lasermeff}{eq:laserham} at $B=0$, \Eqsc{eq:phib}{eq:phibham} at $v/c \ll 1$, and \Eqsd{eq:classphi}{eq:pondham} at $v/c \ll 1$ together with $B = 0$. From \Eq{eq:meffw}, it also readily follows that \mbox{$\meff(a_0<4)>0$}; yet \mbox{$\meff(a_0>4)<0$} at least for some $\sigma>1$ (\Fig{fig:meff}). Other properties of $\meff$ are discussed in \Sec{sec:mult} and C.

\subsection{Tristability}
\label{sec:mult}

\begin{figure}
\centering
\includegraphics[width=0.4 \textwidth]{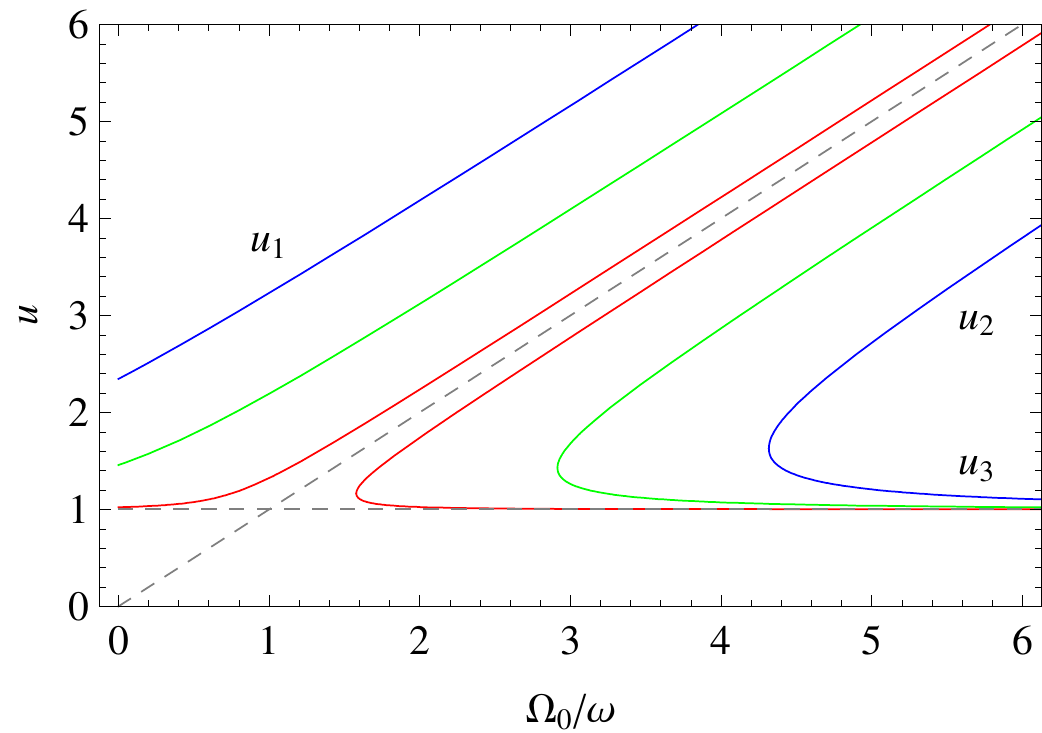}
\caption{(color online) Solution of \Eq{eq:four} for $u$ vs. $\sigma_0 \equiv \Omega_0/\omega$: at $\sigma_0\to \infty$, one has $u_{1,2} \sim \sigma_0 \pm a_0h/\sqrt{2}$, and $u_{3} \sim h\sqrt{1+s^2}$, where $u_i$ correspond to different branches; $u(0) = h\sqrt{1+s^2+a_0^2/2}$. For a given $\sigma_0$, $u_2$ and $u_3$ appear simultaneously behind the nonlinear resonance $u = \sigma_0$ (gray, dashed); the condition is yielded by \Eq{eq:sigmacond}. Shown here is the case $\dv=0$ (so $u = \avr{\gamma}_\xi$), $s=0$: $a_0 = 0.3$ (red), $a_0 = 1.5$ (green), and $a_0 = 3$ (blue). }
\label{fig:uvssigma}
\end{figure}

In the presence of both relativistic effects and nonzero $B$, the cyclotron resonance is essentially nonlinear and permits multiple energy states at given $\dv$ and $\mu$. To see this, rewrite \Eq{eq:up} as
\begin{equation}
\label{eq:four}
\mc{U}(u) \equiv 2(u-\sigma_0)^2(h^{-2}u^2-s^2-1)-a^2_0u^2 = 0,
\end{equation}
where $\sigma_0=\Omega_0/\omega$ \cite{foot:ueq}. \Eq{eq:four} is a fourth-order algebraic equation; thus it allows up to four values of $u$, which also can be found analytically \cite[\S1.8-5]{book:korn}. (Explicit solutions are not shown here because of their complexity.) Since $\mc{U}(0)<0$ and $\mc{U}(\pm \infty)=+\infty$, two solutions always exist, one of them being unphysical ($u_4<0$). Further consideration of the signs in \Eq{eq:four} shows \cite[\S1.6-6(c)]{book:korn} that, apart from degenerate cases, there exist either one or three positive roots,  $u_1>u_2>u_3$. Therefore, one or three energy states are possible (\Fig{fig:uvssigma}), allowing for hysteretic effects \cite{ref:kaplan82, ref:gabrielse85, ref:kaplan85, ref:brown86}, which also have quantum analogies in solid-state physics \cite{ref:levenson67, ref:kaplan84, ref:ashkinadze99}.

\begin{figure*}
\centering
\includegraphics[width=0.95 \textwidth]{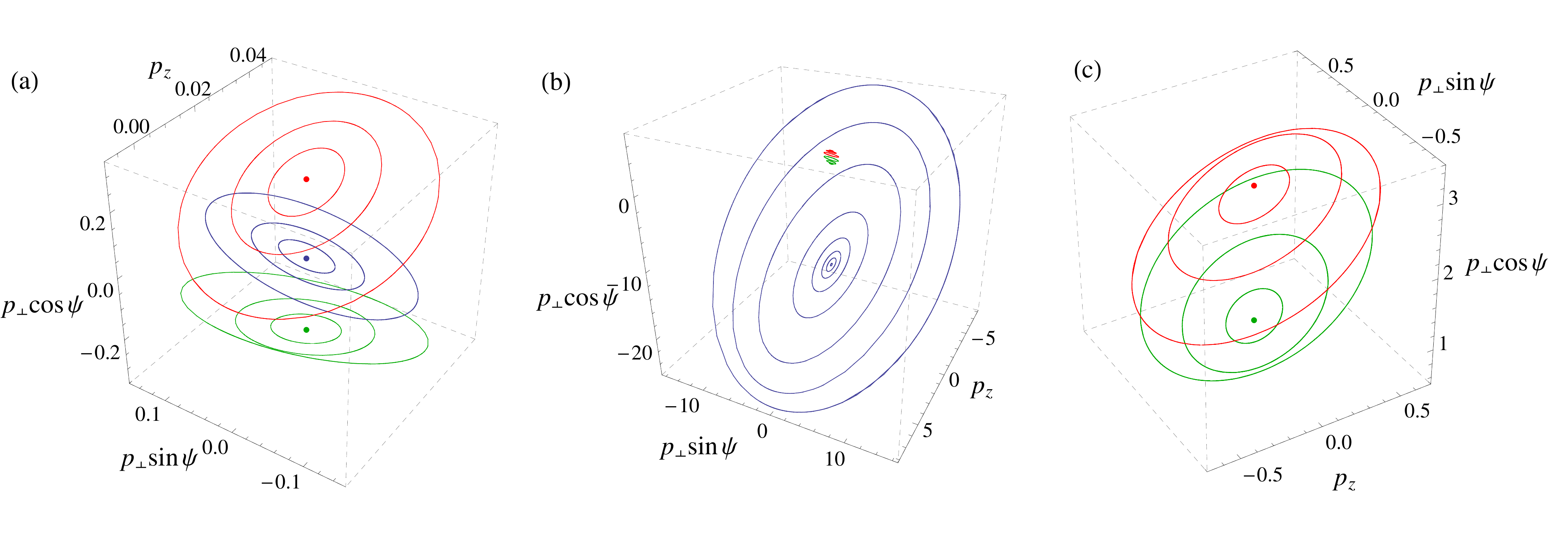}
\caption{(color online) Particle trajectories in the momentum space ($mc$ units) with the same $\rho_0$ but different $s$: (a)~$\rho_0=0.02$, $a_0=10^{-4}\sqrt{2}$, $\sigma_0=1.008$; (b),~(c) $\rho_0=0$, $\sigma_0=8.3$, $a_0=5\sqrt{2}$; (c) close-up of Fig.~b. Blue, red, and green colors denote the branches 1,~2, and 3, correspondingly. All the three equilibria are center-like and result in stable oscillations.}
\label{fig:phaseplot}
\end{figure*}

The condition for multiple $u_i$ reads
\begin{equation}\label{eq:sigmacond}
\sigma_0 > \sigma_c \equiv h\,\Big[(1+s^2)^{1/3}+(a_0^2/2)^{1/3}\Big]^{3/2}.
\end{equation}
Therefore, assuming that three branches exist for a given $s$, there must exist the same number of energy states for $s=0$. The latter energy states correspond to three equilibria in the momentum space $(p_\perp \cos \psi,\,p_\perp \sin \psi,\,p_z)$, where $\psi = \xi+\chi$, $\chi$ being the gyrophase. Apart from the degenerate case when $u_2$ approaches $u_3$, the particle trajectory \eq{eq:xyz} is a continuous function of the initial conditions for each branch. Thus, assuming negligible dissipation (\Sec{sec:lmass}), all the three equilibria are stable here (\Fig{fig:phaseplot}), unlike for a one-dimensional (1D) nonlinear oscillator \cite{book:bogoliubov}, as well as in contrast to the 3D cyclotron resonance in a quasi-static field \cite{ref:brown86} or any wave with a parallel refraction index $n_\subpar$ other than unity \cite{ref:suvorov88, ref:kotelnikov90, ref:litvak93, ref:cohen91}.
 
The difference from \Refs{ref:brown86, ref:suvorov88, ref:kotelnikov90, ref:litvak93, ref:cohen91, ref:nevins87} is understood from the angle-action equations for the transverse oscillations, which now are governed by the Hamiltonian
\begin{equation}\label{eq:1dham}
\mcc{H}=(\sigma-1)\mcc{J}-\lambda\sqrt{\mcc{J}}\,\cos\psi.
\end{equation}
Here $\mcc{J}=\rho_\perp^2/2\sigma$ is the action conjugate to the angle $\psi$, and $\lambda=a_0/\sqrt{\sigma}$; the effective time is $\xi$, $d\xi = (\omega u/\gamma)\,dt$. Since $\rho_z=(1-u^2+2 \sigma\mcc{J})/2u$ [\Eq{eq:xyz}], and $d\vec{r}/d\xi=\vec{p}/m\omega u$, the particle motion stability in the $(\psi,\mcc{J})$ space is equivalent to that in the 6D phase space $(\vec{r},\vec{p})$. (We define $\psi$ through $\chi$ being the cylindrical phase in the momentum space rather than that in the coordinate space. In this case, \Eq{eq:1dham} is exact; otherwise, a similar yet approximate equation follows, in agreement with \Refs{ref:suvorov88, ref:litvak93} for $n_\subpar=1$, $n_\perp=0$, and $v/c \ll 1$.)

For given $\sigma$ and $\lambda(\sigma)$, only one equilibrium, namely, a center, is allowed on the phase plane $(\psi \bmod{2\pi},\mcc{J})$ and located at $\mcc{J}=[\lambda/2(1-\sigma)]^2$, $\psi = 0$ ($\sigma > 1$) or $\psi = \pi$ ($\sigma < 1$). However, as $\sigma=\sigma(u_i)$, there exist up to three different phase planes, and hence the equal number of center points ($i=1,2,3$). This situation is different from that, \eg in \Refs{ref:kotelnikov90, ref:litvak93}, where several equilibria are bound to coexist on a \textit{single} phase plot: as multiple centers are topologically impossible on a plane without a saddle \cite{foot:poincare}, the intermediate-energy equilibrium is unstable and cannot be observed there. For the topological constraint does not apply in our case, all the three equilibria are now stable and equally realized. This results in unusual particle dynamics, which we discuss in \Sec{sec:lmass}.

\subsection{Longitudinal mass}
\label{sec:lmass}

As the three energy states correspond to different effective masses, a guiding center behaves differently depending on which $\meff$ is selected; even the \textit{sign} of the particle acceleration in response to perturbation forces can vary. To see this, rewrite the average motion equation \eq{eq:euler} as
\begin{equation}\label{eq:fpar}
m_\subpar\frac{d\dv}{dt}=F_\subpar,\quad
\end{equation}
where $m_\subpar = \partial\dpp/\partial \dv$, or
\begin{equation}\label{eq:lmass0}
m_\subpar = \frac{\partial}{\partial\dv}
\left(\dgamma\meff\dv-\frac{c^2}{\dgamma}\,\frac{\partial\meff}{\partial\dv}\right),
\end{equation}
is the effective longitudinal mass \cite{ref:okun89, foot:ashcroft},
\begin{equation}
F_\subpar =  -\frac{\partial}{\partial \bar{z}}\,\Bigg[
\dgamma\meff c^2  - \frac{c^2}{\dgamma}\,\left(\dv\,\frac{\partial \meff}{\partial \dv}\right) + e\tilde{\varphi}
\Bigg]- \frac{\partial \dpp}{\partial t}
\end{equation}
is the perturbation force, and $\dpp=\dpp(\bar{z},\dv,t)$ [\Eq{eq:moment1}]. 

\begin{figure}
\centering
\includegraphics[width=0.49 \textwidth]{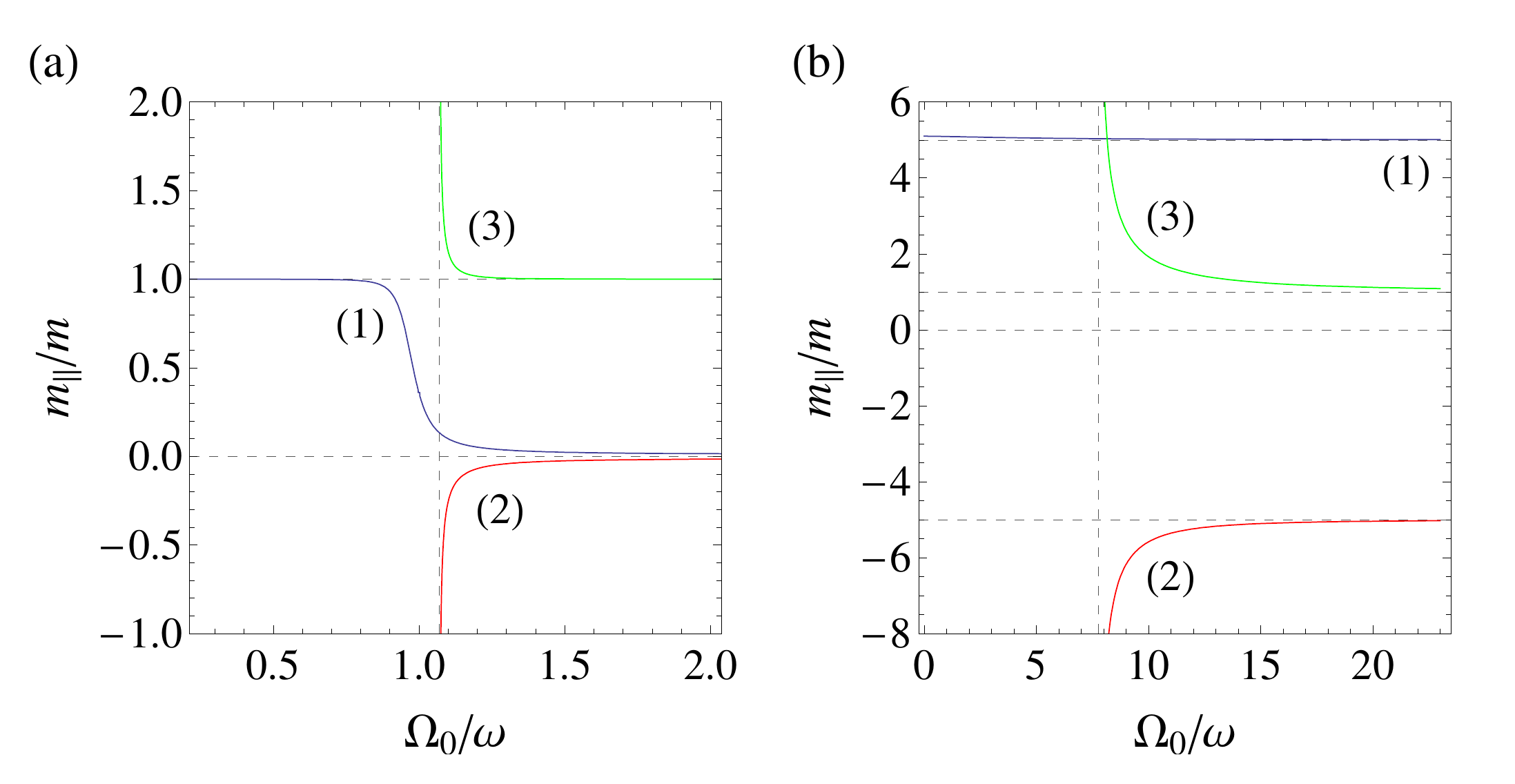}
\caption{(color online) Longitudinal mass $m_\subpar$ in units $m$ vs. $\sigma_0 \equiv \Omega_0/\omega$ for $s=0$, $\dv=0$: (a)~weakly relativistic case, $a_0=10^{-2}\sqrt{2}$; (b)~strongly relativistic case, $a_0=5\sqrt{2}$. The branches corresponding to $u_1$, $u_2$, and $u_3$ are shown in blue, red, and green, respectively; $m_{\subpar 1} (\sigma_0 = 0) = \sqrt{1+a_0^2/2}$. The horizontal dashed lines mark zero and asymptotes at $\sigma_0 \to \infty$: $m_{\subpar 1,2} \to \pm \kern 2pt a_0/\sqrt{2}$, $m_{\subpar 3} \to 1$. The vertical dashed asymptote also marks the transition between the regimes with single and multiple branches; for conditions see the caption of \Fig{fig:uvssigma}.}
\label{fig:mpar}
\end{figure}

A tedious yet straightforward derivation yields
\begin{equation} \label{eq:lmass}
m_\subpar= m\dgamma^3 \frac{\Gamma^{3\kern -.5pt/\kern -.5pt 2}_2}{\Gamma_3}, \quad \Gamma_n=\,1+s^2+\frac{a^2_0}{2(1-\sigma)^n},
\end{equation}
$\Gamma_2$ coinciding with $u'^2$. In the absence of the laser field ($a_0=0$), \Eq{eq:lmass} reads $m_\subpar=\meff\dgamma^3 > 0$ [\Eq{eq:meffb}], as one would expect for a particle with $\meff$ independent of $\dv$ \cite{ref:okun89}. Given a nonzero $a_0$, one as well has $m_{\subpar 1}>0$, because $\sigma<1$ at the first branch, as seen from \Fig{fig:uvssigma}. It is also seen from \Fig{fig:uvssigma} that $u_3 > (\sigma_0/\sigma_c)u_c$, where $u_c \equiv u(\sigma_c)$,
\begin{equation}
u_c = h\, (1+s^2)^{1/3}\Big[(1+s^2)^{1/3}+(a_0^2/2)^{1/3}\Big]^{1/2};
\end{equation}
thus $m_{\subpar 3}>0$, correspondingly. However, $u_2 < (\sigma_0/\sigma_c)\kern 1pt u_c$, yielding $m_{\subpar 2}<0$ for any $\dv$ and $s$ (\Fig{fig:mpar}). 

\begin{figure}
\centering
\includegraphics[width=0.48 \textwidth]{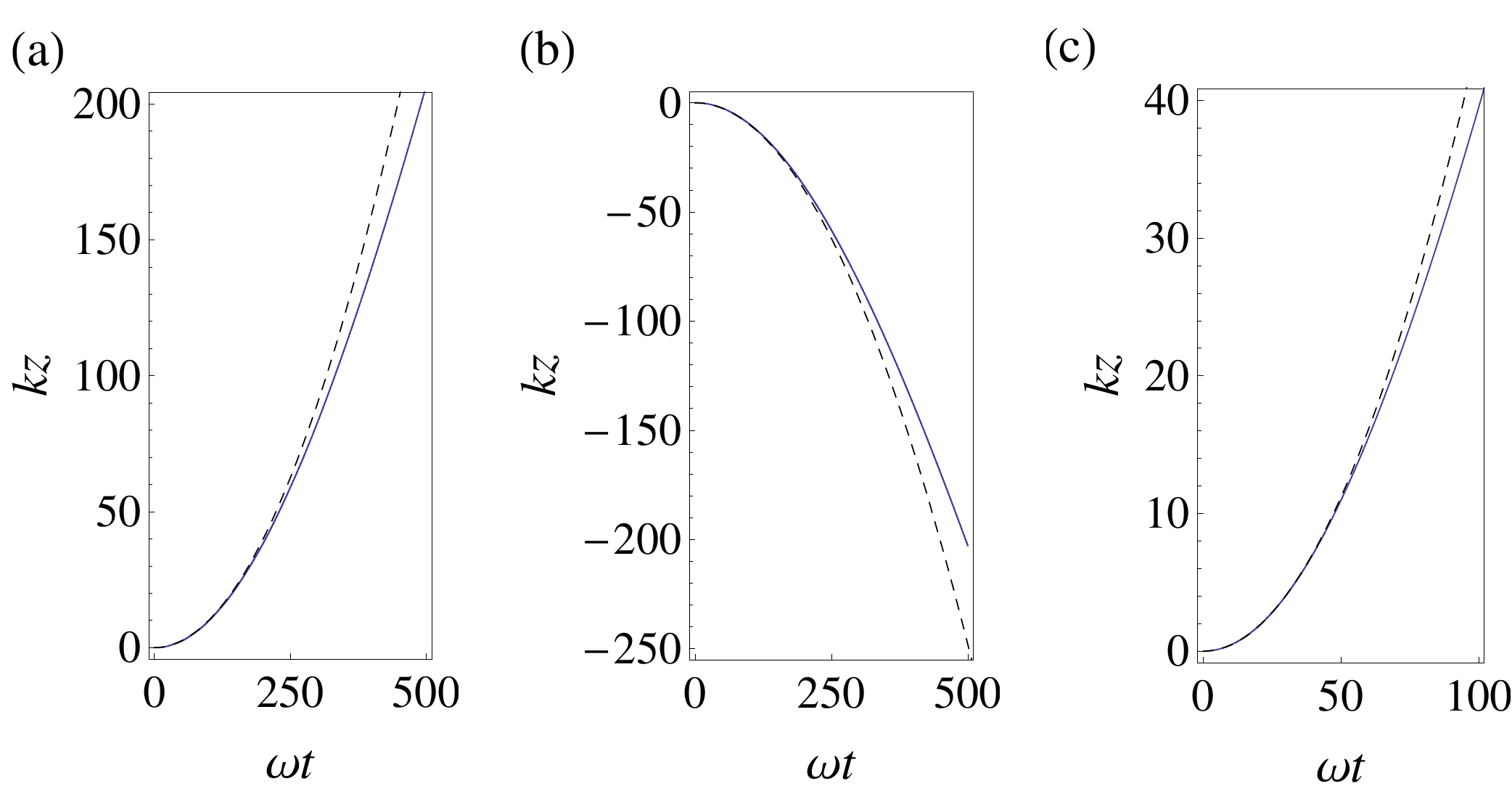}
\caption{$z(t)$ for a particle with initial $\dv=0$, $s=0$ adiabatically accelerated along a magnetic field by a perturbation force $F_\subpar=10^{-2} mc\omega$; $\sigma_0=8.3$, $a_0=5\sqrt{2}$. The sign of acceleration varies depending on the initial energy state 1-3 corresponding to: (a)~$m_{\subpar 1}>0$, (b)~$m_{\subpar 2}<0$, (c)~$m_{\subpar 3}>0$ (see \Fig{fig:mpar}). Solid is the numerical data; dashed are analytic fits $z(t)=F_\subpar t^2/2m_\subpar$, with $m_\subpar$ given by \Eq{eq:lmass}; $z$ and $t$ are measured in units $k^{-1}$ and $\omega^{-1}$, respectively.}
\label{fig:accel}
\end{figure}

\begin{figure}
\centering
\includegraphics[width=0.49 \textwidth]{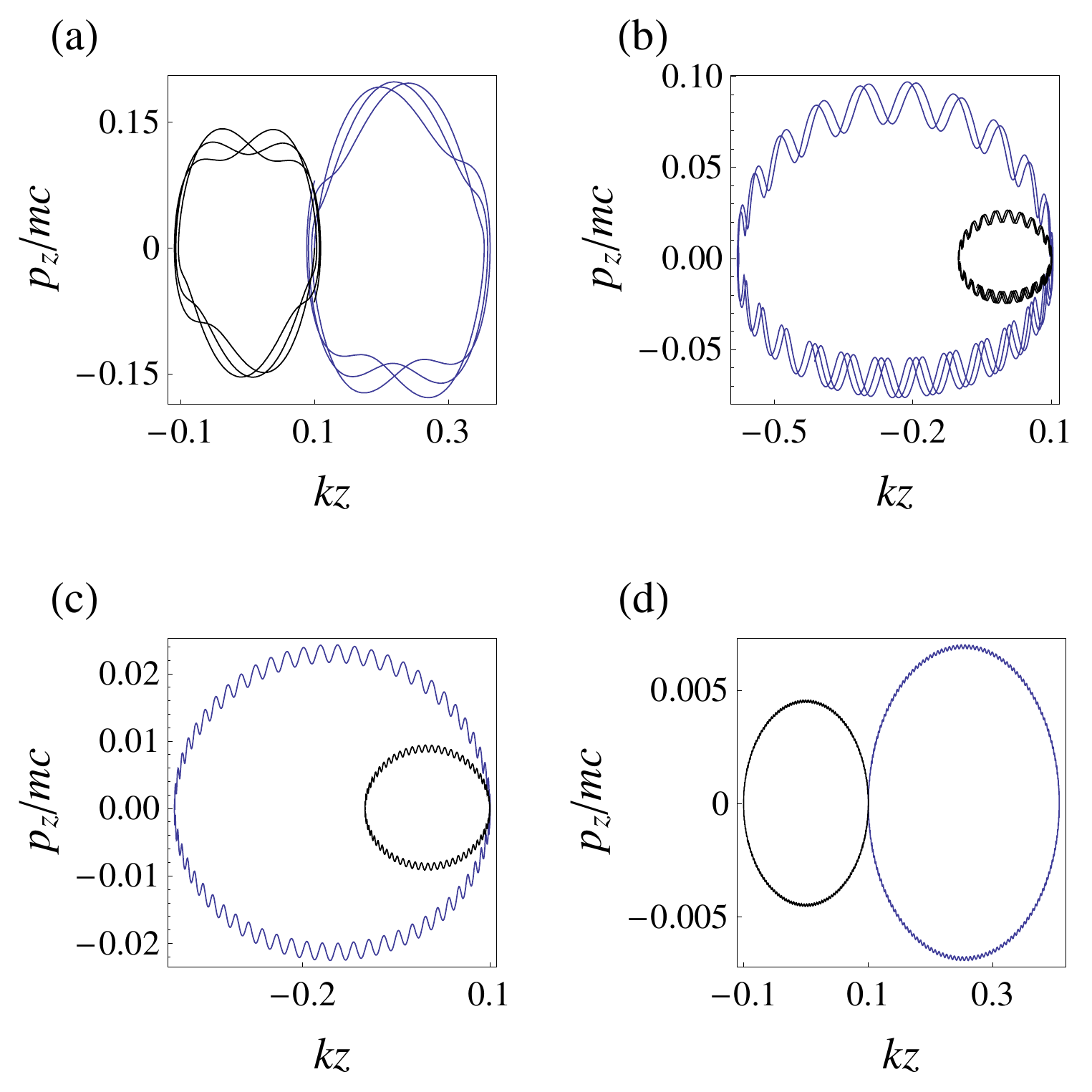}
\caption{(color online) Phase plots $(z,p_z)$ showing bounce oscillations along a magnetic field due to inhomogeneity of $E$ or $B$. An additional perturbation force $F_\subpar$ shifts the equilibrium point in the direction determined by the sign of $F_\subpar/m_\subpar$ (not just $F_\subpar$), with stable bouncing to persist for either sign of $m_\subpar$. (a,\kern 1pt b)~$E(z)$ has a minimum at $z=0$; $B$ is uniform. Bounce oscillations are stable for the branches 1 (Fig.~a) and 2 (Fig.~b) yet unstable for the branch 3 (particles seek high $E$; not shown); $F_\subpar=0$ (black) and $F_\subpar=10^{-2} mc\omega$ (blue). (c,\kern 1pt d)~$E$ is uniform. (c)~$B(z)$ has a maximum at $z=0$. Bounce oscillations are stable for the branch 2 (shown) yet unstable for the branches 1 and 3 (particles seek low $B$; not shown); $F_\subpar=0$ (black) and $F_\subpar=10^{-3} mc\omega$ (blue). (d)~$B(z)$ has a minimum at $z=0$. Bounce oscillations are stable for the branches 1 (not shown) and 3 (shown) yet unstable for the branch 2 (particles seek high $B$; not shown); same $F_\subpar$ as in Fig.~c. All figures: particles are initially at $(0.1,0)$, $s=0$; $z$ and $p_z$ are measured in units $k^{-1}$ and $mc$, respectively. At the extrema: $a_0 \approx 5\sqrt{2}$, and $\sigma_0 \approx 8.3$; $m_{\subpar 1,3}>0$, $m_{\subpar 2}<0$.}
\label{fig:bounce}
\end{figure}

Given the oscillation orbit stability (\Sec{sec:mult}), a particle residing at the second branch will exhibit unusual behavior in response to perturbation forces $F_\subpar$, including gravitational and electrostatic potentials. Unlike a ``normal'' particle with a positive mass, a particle with $m_\subpar < 0$ will accelerate adiabatically in the direction \textit{opposite} to $F_\subpar$ (\Fig{fig:accel}). Alternatively, should the unperturbed particle exhibit bounce oscillations in $z$ (\eg due to inhomogeneity of $E$ or $B$), $F_\subpar$ will shift the equilibrium point in the direction determined by the sign of $F_\subpar/m_\subpar$, with stable bouncing to persist for either sign of $m_\subpar$ (\Fig{fig:bounce}). Merely a dissipative instability is possible for $m_\subpar<0$ (\eg $m_\subpar\dot{\dv}=-\eta \dv$ yields $\dv \propto e^{\eta t/|m_\subpar\kern -1pt|}$); yet it develops on a time scale different from that of the oscillations and, for weak damping, remains insignificant until large $t$.

Transferring particles between the different mass branches also allows for a current drive effect distinguishable from the traditional wave-induced methods, which rely on wave-induced diffusion to higher kinetic energies \cite{ref:fisch87}. The effect is explained as follows. Stationary fields conserve the particle quasi-energy \eq{eq:qenergy} for a given $m_\subpar$ branch and therefore do not permit acceleration along a closed loop. However, should $m_\subpar$ be changed nonadiabatically along the loop, the overall work performed can be nonzero; hence, even curl-free fields such as those due to electrostatic or ponderomotive potentials will be able to produce a continuous energy gain. Similar effects were previously discussed in \Refs{ref:suvorov88, ref:litvak93, my:cdprl, my:cdlarge, my:invited, ref:raizen05, ref:ruschhaupt04, ref:narevicius07, ref:narevicius07b}. With the effective-mass formalism, these effects can now be explained within a unified approach.

\section{Conclusions}
\label{sec:conc}

We showed that a classical particle oscillating in an arbitrary high-frequency or static field effectively exhibits a modified mass $\meff$ derived from the particle averaged Lagrangian [\Eq{eq:meff}]. We obtained relativistic ponderomotive and diamagnetic forces, as well as magnetic drifts, from the $\meff$ dependence on the guiding center location and velocity. The effective mass is not necessarily positive and can result in backward acceleration when an additional perturbation force is applied.

As an example, we explored the average motion of a laser-driven particle immersed in a dc magnetic field. Multiple energy states are realized in this case and yield up to three branches of $\meff$ and the effective longitudinal mass $m_\subpar$ for a given magnetic moment and parallel velocity (\Fig{fig:mpar}). We showed that both $m_\subpar > 0$ and $m_\subpar < 0$ are possible then, the latter regime too allowing for adiabatic dynamics. From other contexts, such negative masses are known to be capable of driving intriguing effects like absolute negative conductivity \cite{ref:elesin05, ref:koulakov03}, negative mass instability \cite{ref:nielsen58, ref:kolomensky59, ref:landau66, ref:lau71, ref:uhm77, ref:bratman95, ref:savilov97, ref:dumbrajs01, ref:strasser02}, and related phenomena \cite{ref:shvartsman94, ref:mehdian01, ref:esmaeilzadeh01}. Yet the effects that may flow from the variable sign of $m_\subpar$ (or $\meff$) particularly for laser-driven particles in a magnetic field remain to be studied.

\section{Acknowledgments}

This work was supported by DOE Contract No. DE-FG02-05ER54838 and by the NNSA under the SSAA Program through DOE Research Grant No. DE-FG52-04NA00139.

\appendix

\section{Guiding center Lagrangian}
\label{sec:dlagr}

Consider a dynamical system, which exhibits slow translational motion in some guiding center variables $(\vec{Q},\vec{P})$ superimposed on fast oscillations in angle-action variables $(\vec{\theta},\vec{J})$. In the adiabatic regime, $\vec{J}$ is conserved, so the system action $S$ can be put in the form 
\begin{equation} 
\label{eq:action1}
S = \vec{J}\cdot\Delta\vec{\theta}+\int \vec{P}\cdot d\vec{Q} - \int H\,dt,
\end{equation}
where $\Delta \vec{\theta}$ is the increment of $\vec{\theta}$ along a given trajectory, $H(\vec{Q},\vec{P};\vec{J})$ is the Hamiltonian, and $t$ is the time. Suppose that we are only interested in guiding center trajectories, that is, those in the $(\vec{Q},\vec{P})$ space. In this case, we can neglect the first term in \Eq{eq:action1}, so as to come up with a reduced variational principle $\delta \mc{S}=0$, where
\begin{equation} 
\label{eq:action2}
\mc{S} = \int \left(\vec{P}\cdot \dot{\vec{Q}} - H\right)\,dt
\end{equation}
is the new action to be varied with respect to $\vec{Q}$ and $\vec{P}$ only (cf., \eg Ref.~\cite[\S44]{book:landau1}). 

Using $S=\int L\,dt$, where $L$ is the Lagrangian, \Eq{eq:action2} can be written as \cite{foot:routhian}
\begin{equation} 
\label{eq:action3}
\mc{S} = \int \left(L- \vec{J}\cdot\dvectheta\kern .5pt\right)\,dt.
\end{equation}
By definition, the integrand here must be expressed in terms of the guiding center variables only. Hence $\dvectheta = \vec{\nu}(\vec{Q},\vec{P})$ (a parametric dependence on $\vec{J}$ is implied hereafter), so $\vec{J}\cdot\dvectheta\,dt$ is \textit{not} an exact differential, and the first term in \Eq{eq:action3} is transformed as follows. By definition, $L$ is a periodic function of $\vec{\theta}$, except that it may contain nonperiodic terms that are full time derivatives. Since omitting the latter does not affect the motion equations, below we assume that $L = \avr{L}+L_\sim$, where the angle brackets stand for time averaging, $\avr{L_\sim} = 0$, and $\avr{L}$ is a function of $(\vec{Q},\vec{P})$, or $(\vec{Q},\dot{\vec{Q}})$ only. On time scales of interest, that is, $\Delta t \gg  \nu_i^{-1}$ and $\Delta t \gg |\nu_i-\nu_j|^{-1}$ ($\nu_i$ being any of the oscillation frequencies), the oscillatory term in \Eq{eq:action3} vanishes; thus
\begin{equation} 
\label{eq:action5}
\mc{S} = \int \Big[\!\avr{L}-\vec{J}\cdot\vec{\nu}\Big]\,dt.
\end{equation}
We can now introduce a guiding center Lagrangian $\mc{L}(\vec{Q},\dot{\vec{Q}})$ as $\mc{S}=\int \mc{L}\,dt$. Since the equality
\begin{equation} 
\label{eq:action6}
\int \mc{L}\,dt = \int \Big[\!\avr{L}-\vec{J}\cdot\vec{\nu}\Big]\,dt
\end{equation}
must hold for any time interval, one has
\begin{equation} 
\label{eq:lagr1}
\mc{L} = \avr{L}-\vec{J}\cdot\vec{\nu},
\end{equation}
in agreement with \Refs{ref:larsson86, ref:pfirsch04}. (For a system exhibiting oscillations on multiple time scales, different $\mc{L}$s can be introduced depending on how the time averaging is defined.) One can also show that \Eq{eq:lagr1} conforms to the requirement of gauge invariance: Replacing $L$ with $L+dG/dt$, where $G(\vec{\theta},\vec{Q},t)$ is an arbitrary function, will result in $\mc{L} \to \mc{L} + \Delta \mc{L}$, where
\begin{equation} 
\Delta \mc{L} = \vec{\nu}\cdot\avr{\frac{\partial G}{\partial \vec{\theta}}}+\dot{\kern -2pt\vec{Q}}\cdot \frac{\partial \bar{G}}{\partial \vec{Q}}+\frac{\partial \bar{G}}{\partial t},
\end{equation}
and $\bar{G}(\vec{Q},t)=\avr{G(\vec{\theta},\vec{Q},t)}$. The first term is equal to zero due to $G$ being periodic in $\vec{\theta}$. Thus, one has $\Delta \mc{L} = d\bar{G}/dt$, \ie $\Delta \mc{L}$ is, too, a full time derivative, and therefore does not affect the average motion equations
\begin{equation} 
\label{eq:euler}
\frac{d}{dt}\left(\frac{\partial \mc{L}}{\partial \dot{\vec{Q}}}\right)=\frac{\partial \mc{L}}{\partial \vec{Q}}.
\end{equation}

For $L$ being a periodic function of $t$ (rather than, or in addition to, $\vec{\theta}$), the above procedure would map out the time variable, thus yielding an analogue of the Maupertuis principle \cite[\S44]{book:landau1}. However, should $t$ be kept (unlike $\vec{\theta}$) as the independent variable, the derivation of the guiding center Lagrangian is modified as follows. Consider the fast time $\tilde t$ and the slow time $\hat{t}$ separately. Then we obtain an extended system having the action
\begin{equation} 
\label{eq:action7}
\hat{S} = \vec{J}\cdot\Delta\vec{\theta} -\tilde{H}\,\Delta\tilde{t} 
+ \int \vec{P}\cdot d\vec{Q} - \int\hat{H}\,d\hat{t},
\end{equation}
where the formally introduced momentum $-\tilde{H}$ conjugate to $\tilde{t}$ is to remain constant in the adiabatic regime. The super-Hamiltonian $\hat{H}$ must generate the same canonical equations as those of the original system; it must also provide that $d\tilde{t}/d\hat{t} = 1$, as follows from the definition of $\tilde{t}$. These conditions are satisfied if one takes $\hat{H}=H + \tilde{H}$, so the super-Lagrangian 
\begin{equation} 
\label{eq:lagr2}
\hat{L} = \vec{P}\cdot\dot{\vec{Q}}+\vec{J}\cdot\dvectheta - \tilde{H}\dot{\tilde{t}}-\hat{H}
\end{equation}
equals $L$. Then the guiding center Lagrangian reads
\begin{equation} 
\label{eq:lagr3}
\hat{\mc{L}} = \avr{L}-\vec{J}\cdot\vec{\nu}+\tilde{H},
\end{equation}
which is equivalent to \Eq{eq:lagr1}, since constant $\tilde{H}$ can be omitted.

Results similar to those in this Appendix were obtained earlier for particle motion in a dc magnetic field \cite{ref:danilkin95, ref:beklemishev99, ref:brizard07}, oscillations in nonrelativistic high-frequency waves \cite{my:dipole, my:invar}, and laser-driven relativistic electron dynamics in vacuum \cite{ref:bauer95, my:meff, ref:moore94}. In the main text, we make use of the general form of the theorem \eq{eq:lagr1}, which contains the earlier results as particular cases. This generality allows us to formulate a fundamental concept of the effective mass for an oscillating particle without making preliminary assumptions on the nature of the oscillations.

\section{Ponderomotive forces}
\label{sec:ponder}

\subsection{Nonrelativistic wave fields}

Let us apply the effective mass formalism to derive ponderomotive forces, starting with those in the nonrelativistic regime \cite{ref:gaponov58, ref:cary77, ref:motz67, my:tunnel, my:invited}. Consider a particle oscillating in a high-frequency wave $\vec{E}=-\nabla \varphi$, where
\begin{equation}
\varphi = \varphi_0(\vec{r},t)\cos(\omega t - \vec{k}\cdot\vec{r}).
\end{equation}
We will assume that $k \dv \ll \omega$; we will also assume that the envelope $\varphi_0(\vec{r},t)$ varies little on the time scale $\omega^{-1}$ and has a spatial scale $\ell$ large compared to the amplitude of the particle oscillations ($\ell \gg eE/m\omega^2$) and the guiding center displacement on the oscillation period ($\ell \gg \dv/\omega$) \cite{my:tunnel}. Then $\mc{L}'=-mc^2+\avr{L_\text{osc}}$, where
\begin{gather}
L_\text{osc} \approx \frac{m\dot{\vec{r}}^2_\text{osc}}{2}- e\kern 1pt\vec{r}_\text{osc} \cdot \vec{E}(\dvr,t)
\end{gather}
is obtained using $\varphi(\vec{r}) \approx \varphi({\dvr})-\vec{r}_\text{osc} \cdot \vec{E}$, with the quiver displacement $\vec{r}_\text{osc}=-e\vec{E}/m\omega'^2$, and the Doppler-shifted frequency $\omega' = \omega - \vec{k}\cdot\dvv$. Then $\avr{L_\text{osc}}= -\Phi$, where
\begin{equation}
\label{eq:classphi}
\Phi = \frac{e^2E^2_0}{4m(\omega - \vec{k}\cdot\dvv)^2}
\end{equation}
is known as the ponderomotive potential \cite{ref:gaponov58, ref:cary77, ref:motz67, my:tunnel, my:invited}, $E_0$ being the field amplitude; thus
\begin{equation}
\meff = m + \Phi/c^2.
\end{equation}
Omitting an insignificant constant, the guiding-center Lagrangian reads $\mc{L}=\frac{1}{2}\,m\dv^2 - \Phi$. Hence, the Hamiltonian takes the well known form
\begin{equation}
\label{eq:pondham}
\mc{H} = \frac{1}{2m}\,\dpp^2+\Phi,
\end{equation}
$\Phi$ playing a role of an effective potential, as expected.

Suppose now that, under the same conditions, an additional dc magnetic field $\vec{B}$ is imposed. Assuming that $\vec{B}$ is smooth, one has $\mc{L}'=-mc^2-\mu B+\avr{L_\text{osc}}$, where
\begin{equation}
\label{eq:murf}
\mu = \frac{m}{2B}(\vec{v}_\perp - \vec{v}_\text{osc})^2
\end{equation}
is the new adiabatic invariant proportional to the action of the particle Larmor rotation at frequency $\Omega_0=eB/mc$, $\vec{v}_\text{osc}$ is the induced oscillatory velocity proportional to $E$ \cite{my:invar, ref:motz67, ref:watson68, ref:eubank69}, and $\avr{L_\text{osc}}$ is proportional to $E^2$. Suppose $\vec{B} \approx \unitvec{z}B(z)$ [\Eq{eq:ab}] and take
\begin{equation}
\vec{E}={\rm Re}\left( E_0^\supplus\unitvec{\tau}^\supplus + E_0^\supminus\unitvec{\tau}^\supminus + E_0^\suppar\unitvec{\tau}^\suppar\right) e^{-i\omega t+ikz},
\end{equation}
where $E_0^{(j)}$ are smooth envelopes, and $\unitvec{\tau}^{(j)}$ are unit vectors denoting polarization circular in the plane transverse to the dc magnetic field and that parallel to $\vec{B}$:
\begin{equation}
\unitvec{\tau}^\suppm=(\unitvec{x} \pm i \unitvec{y})/\sqrt{2}, \qquad \unitvec{\tau}^\suppar_{0}=\unitvec{z}.
\end{equation}
In this case, $\avr{L_\text{osc}} = -\Phi_B$ \cite{my:invar}, where
\begin{equation}
\label{eq:phib}
\Phi_B = \frac{e^2}{4m\omega'^2}\Bigg\{ \frac{\big|E_0^\supplus\kern -1pt\big|^2}{1+\Omega_0/\omega'}+\frac{\big|E_0^\supminus\kern -1pt\big|^2}{1-\Omega_0/\omega'}+ \big|E_0^\suppar\kern -1pt\big|^2 \Bigg\}
\end{equation}
matches the known ponderomotive potential in a dc magnetic field \cite{ref:gaponov58, ref:motz67}. Thus, $\meff = m + (\mu B + \Phi_B)/c^2$, and 
\begin{equation}
\label{eq:phibham}
\mc{H} = \frac{1}{2m}\,\dpp^2+\mu B+\Phi_B,
\end{equation}
in agreement with the earlier results \cite{my:invar}.

\mbox{}

\subsection{Relativistic laser wave in vacuum}

Now consider relativistic electron motion in a vacuum laser field $\vec{A}= \vec{A}_0(\vec{r},t)\cos \xi$ of arbitrary polarization, assuming that the vector-potential envelope $\vec{A}_0(\vec{r},t)$ has a scale large compared to the wavelength, $\xi = \omega(t-\vec{n}\cdot \vec{r}/c)$ is the phase, and $\unitvec{n}=\vec{k}/k$ is a unit vector, say, in the $\unitvec{z}$ direction. Using \Eqsd{eq:ucons}{eq:dvvrel} and conservation of the transverse canonical momentum $\vec{p}_\perp' =- (e/c)\vec{A}'$, one has
\begin{equation}
\mc{L}'=-mc^2\frac{1+\avaprimesq_\xi}{\avr{\gamma'}_\xi},
\end{equation}
where $\vec{a}' = e\vec{A}'/mc^2$, and $\avaprimesq_\xi = \favr{a^2}_\xi$ is an invariant. Using \Eq{eq:up1}, one also has, without solving the motion equations, that
\begin{gather}
\avr{\gamma'}_\xi = \sqrt{1+\favr{\ppprime^2}_\xi\!/(mc)^2} = \sqrt{1 + \avr{a^2}_\xi}, 
\end{gather}
and $\mc{L}'=-mc^2\avr{\gamma'}_\xi$. Thus, $\meff$ equals \cite{ref:kibble66, ref:eberly68, ref:sarachik70, ref:rax92, ref:rax92b, ref:rax93, ref:rax93b, ref:moore94, ref:bauer95, ref:mora97, ref:tokman99, my:meff, my:gev, ref:bourdier05b}
\begin{equation}
\label{eq:lasermeff}
\meff=m\sqrt{1+\avr{a^2}_\xi},
\end{equation}
which is a Lorentz invariant independent of $\dvv$. Correspondingly, the guiding center momentum reads $\bar{\vec{p}} = \dgamma \meff\dvv$, and the well known Hamiltonian is given by
\begin{equation}
\label{eq:laserham}
\mc{H} = \sqrt{\meff^2c^4 + \dpp^2c^2},
\end{equation}
the ponderomotive force resulting from the $\meff$ dependence on $\dvr$ and, possibly, slow dependence on $t$.


\begin{thebibliography}{100}
\bibitem{foot:quant}
In nonrelativistic quantum mechanics, a similar ``dressed-particle''
  formulation is used; see, \eg \Refs{ref:dalibard85, ref:guo92, ref:guo92b}.
\bibitem{foot:dressed}
Nonoscillatory interactions are also known to act as modifications of the
  particle properties, \eg in classical plasma physics \cite{book:ichimaru,
  ref:vollick95}, solid state physics \cite[Chap.~12]{book:ashcroft}, and quantum field
  theory \cite{ref:greenberg58}.
\bibitem{foot:jackson}
The argument followed here is similar to that used for free particles; see, \eg
  \Ref{book:jackson} or \Ref{book:landau2}.
\bibitem{foot:routhian}
The quantity $\vec{J}\cdot\vec{\nu}-L$ is known as the Routhian
  \cite[\S41]{book:landau1} for changing to canonical variables
  $(\vec{\theta},\vec{J})$, where $\dvectheta=\vec{\nu}$.
\bibitem{foot:ueq}
The velocity dependence can be removed from \Eq{eq:four} by the variable
  transformation $u=u'h$, $\sigma_0=\sigma_0'h$.
\bibitem{foot:poincare}
Finite motion is assumed, so the Poincar\'e index theorem applies; see, \eg
  \Ref{ref:smith84}.
\end{thebibliography}
\end{document}